\documentstyle[epsf]{article}    
\def\title#1{\centerline{\Large \bf #1}\kern10.pt}
\def\subtitle#1{\centerline{\Large \bf #1}\kern10.pt}
\def\author#1{\centerline{#1}}
\def\address#1{\centerline{\em #1}}

\begin{document}
\small
%
%
\title{The Mystery of the Ramsey Fringe}
\subtitle{that Didn't Chirp}
\author{D.~M. Harber, H.~J. Lewandowski, J.~M. McGuirk\cite{qpdNIST}, and E.~A. Cornell\cite{qpdNIST}}
\address{JILA, National Institute of Standards and Technology and Department of
Physics,}
\address{University of Colorado, Boulder, Colorado 80309-0440}
\vskip10.pt

\centerline{\bf Abstract} We use precision microwave spectroscopy
of magnetically trapped, ultra-cold $^{87}$Rb to characterize
intra- and inter-state density correlations.  The cold collision
shifts for both normal and condensed clouds are measured.  The
results verify the presence of the sometimes controversial
``factors of two'', in normal-cloud mean-field energies, both within
a particular state and between two distinct spin species.  One
might expect that as two spin species decohere, the inter-state
factor of two would revert to unity, but the associated frequency
chirp one naively expects from such a trend is not observed in our
data.

\section{\bf Introduction}
When one studies the statistics of arrival times of photons
emitted from an incoherent source, one finds that immediately
after detecting a photon, one is twice as likely to detect a
second photon than one would expect from the time-averaged
detection rate.  This ``photon bunching'' effect can be understood
as arising naturally from the quantum statistics of a noisy
bosonic field.  If one detects a photon, one can infer that the
randomly fluctuating boson field is near a peak in its amplitude.
Small wonder then that one is likely to detect another photon
soon.  Similar effects are seen in the correlations in the arrival
times of bosonic atoms falling from a cold cloud \cite{shimizu}.
Within an atomic cloud itself, these statistical effects are best
thought of as density fluctuations.  Just as the short-time peak
in photon arrival statistics is suppressed in a coherent (laser)
beam of photons, the density fluctuations in a cloud of bosonic
atoms is suppressed if the atoms are Bose-condensed.  This effect
has been seen in the analysis of expansion energy of condensates
\cite{holland1997,ketterle1997} and in the comparison of
three-body recombination rates in condensates versus thermal clouds
\cite{jila1997}.

In a recent series of experiments we examined the effects of these
density fluctuations on the hyperfine transition frequency in
ultra-cold normal and in Bose-condensed rubidium \cite{icap}. The
MIT hydrogen group performed early work in this area
\cite{Hcoldc}. In this shorter conference proceedings, we review
our spectroscopic study \cite{icap} with particular emphasis on
the effects of decoherence on the density correlations between two
distinct hyperfine states. A cloud of thermal atoms begins
initially in a single hyperfine state.  A microwave pulse
coherently splits the cloud into two distinct hyperfine states
with a well-defined relative phase. Ramsey spectroscopy, sensitive
to the density effects, shows that the two states initially have
the same factor of two in their inter-state density correlations
as each does in its intra-state density correlations.  What we
find counter-intuitive is that as time evolves and the two spin
species begin to decohere, we see no corresponding shift in the
frequency of the Ramsey fringes.  Thus, with apologies to Arthur
Conan Doyle, we came up with the title of the present manuscript.

\section{\bf Hyperfine Spectroscopy}

Spatial inhomogeneity of the transition frequency was minimized
through the use of a pair of energy levels which experience the
same trapping potential.  At a magnetic field of $\sim3.23$ G the
$|1\rangle\equiv|F=1,m_{f}=-1\rangle$ and
$|2\rangle\equiv|F=2,m_{f}=1\rangle$ hyperfine levels of the
$5\mbox{S}_{1/2}$ ground state of $^{87}$Rb experience the same
first-order Zeeman shift.  For a normal cloud at 500 nK, each
energy level is Zeeman shifted by $\sim$10 kHz across the extent
of the cloud, however at 3.23 G the \textit{differential} shift of
the two levels across the cloud is $\sim$1 Hz.  Compared to the
differential Zeeman shift, the energy shift due to cold collisions
is then a relatively large effect at high densities, making
measurements of collisional shifts in this system possible.  The
small inhomogeneity allows for long coherence times, $\sim$2
seconds and longer for low-density clouds, making this system
attractive for precision measurements as well as for the study of
coherence in finite temperature systems.

The experimental setup has been previously described
\cite{lewan2002} and will be briefly summarized here.
Approximately $10^{9}$ $^{87}$Rb atoms are loaded into a vapor
cell magneto-optical trap (MOT).  The atoms are then optically
pumped into the $|F=1\rangle$ state by turning off the repump beam
while the MOT beams remain on.  Then the trapping beams are turned off
and the MOT coils are ramped to a high current, forming a 250 G/cm
gradient to trap $|1,-1\rangle$ atoms in the quadrupole field of
the coils.  The quadrupole coils are mounted on a linear
servo-motor controlled track which then moves the coils 44 cm,
from the MOT region to a Ioffe-Pritchard trap in the ultra-high
vacuum region of the system.  The Ioffe-Pritchard trap consists of
two permanent magnets which provide a 450 G/cm radial gradient.
Two pairs of electromagnetic coils, a pinch and a bias, provide
confinement in the axial direction, which is aligned perpendicular
with respect to gravity.  At a typical bias field of 3.23 G atoms
in the $|1,-1\rangle$ state experience $\{230,230,7\}$ Hz trap
frequencies.  The sample is further cooled by rf evaporation, and
condensates of up to $10^{6}$ atoms can be formed.  Imaging is
performed by the use of adiabatic rapid passage to transfer atoms
from the $|1,-1\rangle$ state to the $|2,-2\rangle$ state.
Anti-trapped $|2,-2\rangle$ atoms rapidly expand for 2-5 ms and
then are imaged through absorption by a 20 $\mu\mbox{s}$ pulse of
$5\mbox{S}_{1/2}$ $|2,-2\rangle \rightarrow 5\mbox{P}_{3/2}$
$|3,-3\rangle$ light.

A two-photon microwave-rf transition is used to transfer atoms
between the $|1\rangle$ and $|2\rangle$ states.  A  detuning of
0.7 MHz from the $|2,0\rangle$ intermediate state provides a
two-photon Rabi frequency of $\sim2.5$ kHz.  Ramsey spectroscopy
of the $|1\rangle \rightarrow |2\rangle$ transition is performed
by measuring the total number of atoms remaining in state
$|1\rangle$ after a pair of $\frac{\pi}{2}$ pulses separated by a
variable time delay is applied \cite{ramsey1956}.  The frequency
of the resulting Ramsey fringes is the difference between the
transition frequency $\nu_{12}$ and the two-photon drive
frequency.  In previous work we measured local variations of
$\nu_{12}$  by detecting the number of atoms remaining in state
$|1\rangle$ at specific spatial locations along the axis of the
normal cloud \cite{lewan2002}.  By analyzing the spatio-temporal
variations of $\nu_{12}$, combined with the measured evolution of
the $|1\rangle$ state after a single $\frac{\pi}{2}$ pulse, we
were able to spatially resolve the evolution of spin waves
\cite{mcguirk2002}.  In this work, in order to perform
measurements of $\nu_{12}$ insensitive to spin waves, one of the
following two techniques was used.  With one technique the entire
cloud, rather than specific spatial locations, was monitored to
average out the effects of spin waves.  Alternatively Ramsey
spectroscopy was restricted to interrogation times short compared
to the spin wave frequency \cite{fspinwave}.

One effect which shifts the transition frequency $\nu_{12}$ is the
differential Zeeman shift.  The Breit-Rabi formula predicts a
minimum in $\nu_{12}$ at $B_{0}=3.228 917(3)$ Gauss, thus the
$|1\rangle$ and $|2\rangle$ energy levels experience an identical
Zeeman shift at $B=B_{0}$.  The differential Zeeman shift about
$B_{0}$ can be approximated as $\nu_{12} = \nu_{min}+\beta
(B-B_{0})^{2}$ \cite{brettrabi}.  Measuring $\nu_{12}$ for
different magnetic fields allows us to calibrate our magnetic
field from the expected dependence.  By working at the vicinity of
$B=B_{0}$ we greatly reduce spatial inhomogeneity of $\nu_{12}$
and also become first-order insensitive to temporal magnetic field
fluctuations.

\section{\bf Density Shifts}

A second effect which shifts $\nu_{12}$ arises from atom-atom
interactions.  In the s-wave regime, where the thermal de Broglie
wavelength of the atoms is greater than their scattering length, atoms experience an energy shift equal to
$\alpha\frac{4\pi\hbar^{2}}{m}an$, where $\alpha$ is the
two-particle correlation at zero separation, $n$ is atom number
density, $a$ is the scattering length, and $m$ is the atom mass.
Therefore for a two-component sample the expected energy shift of
each state is

\begin{eqnarray}
\delta \mu_{1} = \frac{4\pi\hbar^{2}}{m}(\alpha_{11}a_{11}n_{1} +
\alpha_{12}a_{12}n_{2})\\ \delta \mu_{2} =
\frac{4\pi\hbar^{2}}{m}(\alpha_{12}a_{12}n_{1} +
\alpha_{22}a_{22}n_{2}),
\end{eqnarray}

\noindent where $n_1$ and $n_2$ are the $|1\rangle$ and
$|2\rangle$ state densities and

\begin{equation}
\alpha_{ij} =
\frac{\langle\Psi^{\dag}_{i}\Psi^{\dag}_{j}\Psi_{i}\Psi_{j}\rangle}{\langle\Psi^{\dag}_{i}\Psi_{i}\rangle\langle\Psi^{\dag}_{j}\Psi_{j}\rangle}\mbox{.}
\end{equation}

\noindent The shift of the transition frequency in Hz can then be
written as

\begin{eqnarray}
\Delta \nu_{12} & = & (\delta \mu_2-\delta \mu_1)/h \nonumber\\
& = & \frac{2\hbar}{m}(\alpha_{12}a_{12}n_{1} +
\alpha_{22}a_{22}n_{2}-\alpha_{11}a_{11}n_{1} -
\alpha_{12}a_{12}n_{2}) \nonumber\\
 & = & \frac{\hbar}{m}n(\alpha_{22}a_{22}-\alpha_{11}a_{11} + \nonumber\\
 &   & (2\alpha_{12}a_{12}-\alpha_{11}a_{11}-\alpha_{22}a_{22})f)
\end{eqnarray}

\noindent where

\begin{equation}
f=\frac{n_1-n_2}{n}
\end{equation}
and $n=n_1+n_2$.

For non-condensed, indistinguishable bosons, $\alpha=2$ due to
exchange symmetry, therefore $\alpha_{11}^{nc}=\alpha_{22}^{nc}=2$
in a cold normal cloud (where the superscript $c$ or $nc$ refers
to condensed of non-condensed atoms respectively).
Distinguishable particles do not maintain exchange symmetry,
making $\alpha_{12}^{nc}=1$ for a incoherent two-component
mixture.  However if a two-component sample is prepared by
coherently transferring atoms from a single component, such as in
Ramsey spectroscopy, then the excitation process maintains
exchange symmetry, and we might expect $\alpha_{12}^{nc}=2$
\cite{Kleppner2002}.  In this scenario the collisional shift
should be calculated using
$\alpha_{11}^{nc}=\alpha_{22}^{nc}=\alpha_{12}^{nc}=2$, leading to
a predicted frequency shift of

\begin{equation}
\Delta \nu_{12} = \frac{2\hbar}{m}n(a_{22}-a_{11} +
(2a_{12}-a_{11}-a_{22})f).
\end{equation}

\begin{figure}
\leavevmode \epsfxsize=3.375in \epsffile{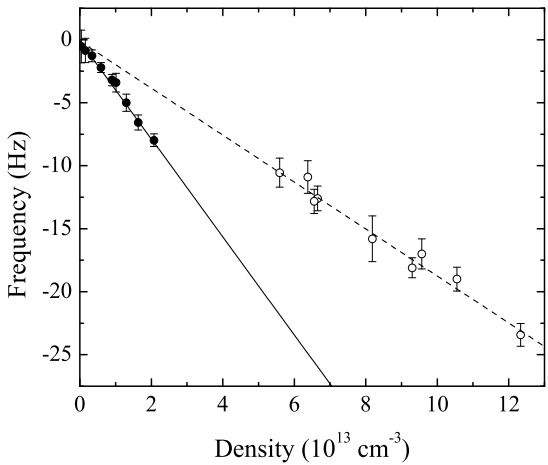}
\caption{\label{fig:dshift}  Measurement of the cold collision
shift.  Solid and open circles represent measurements of the
normal cloud and condensate respectively.  The solid line is a fit
to the normal cloud data $\Delta \nu_{12} = 0.1(0.4) - 3.9(0.3)
10^{-13} n$; the dashed line is a fit to the condensate data
$\Delta \nu_{12} = -0.1(1.4) - 1.9(0.2) 10^{-13} n$ where $\Delta
\nu_{12}$ is in Hz and $n$ is in cm$^{-3}$.}
\end{figure}

\noindent This result can also be obtained by solving the
transport equation \cite{levitov2002,williams2002}.   From
spectroscopic studies \cite{verhaar2002} the three $^{87}$Rb
scattering lengths of interest have been determined to be
$a_{22}=95.47a_0$, $a_{12}=98.09a_0$, and $a_{11}=100.44a_0$,
where $a_0$ is the Bohr radius.  The frequency shift can then be
written as

\begin{equation}
\Delta \nu_{12}= \frac{2\hbar}{m}a_{0}n(-4.97+0.27f).
\end{equation}

\noindent If on the other hand the $|1\rangle$ and $|2\rangle$
states do \textit{not} maintain exchange symmetry, such that
$\alpha_{12}^{nc}=1$, then the frequency shift would instead be

\begin{equation}
\Delta \nu_{12}= \frac{2\hbar }{m}a_{0}n(-4.97-97.82f).
\end{equation}

\noindent These two models are clearly distinguished by the
dependence of $\nu_{12}$ on $f$.

When we perform Ramsey spectroscopy with a pair of $\frac{\pi}{2}$
pulses, the populations of the $|1\rangle$ and $|2\rangle$ states
are equal, and thus $f=0$ during the interrogation time.  From
Eq.~(4) it is apparent that with $f=0$ the collisional shift is
sensitive only to $\alpha^{nc}_{ii}$ and $a_{ii}$ terms.  For
these measurements the bias field was set to $B_0$, and the
transition frequency was measured for a range of densities.  To
adjust density of the sample, the number of atoms in the initial
MOT load was varied.  All normal cloud data was taken at the same
temperature of 480 nK, and all condensate data was taken with high
condensate fractions in order to minimize effects due to the
normal cloud.  The density for the normal cloud was found by
fitting Gaussian profiles to absorption images of the clouds and
extracting the number, temperature, and density.  To measure
condensate density Thomas-Fermi profiles were fit to absorption
images of the condensates and the total number, $N_0$, in the
condensates and the Thomas-Fermi radius along the long axis, $Z$,
were extracted.

The results of this measurement are shown in
Fig.~\ref{fig:dshift}.  Comparing the collisional shift measured
for the normal cloud to that measured for a condensate gives
$\alpha^{nc}_{ii}/\alpha^{c}_{ii}=2.1(2)$.  If instead we
\textit{assume} $\alpha^{nc}_{ii}=2$ and $\alpha^{c}_{ii}=1$, then
the data for both the condensate and normal cloud can be used to
obtain a value for the difference in scattering lengths of
$a_{22}-a_{11}=-4.92(28)a_0$, in agreement with values determined
from molecular spectroscopy \cite{verhaar2002}.

\begin{figure}
\leavevmode \epsfxsize=3.375in \epsffile{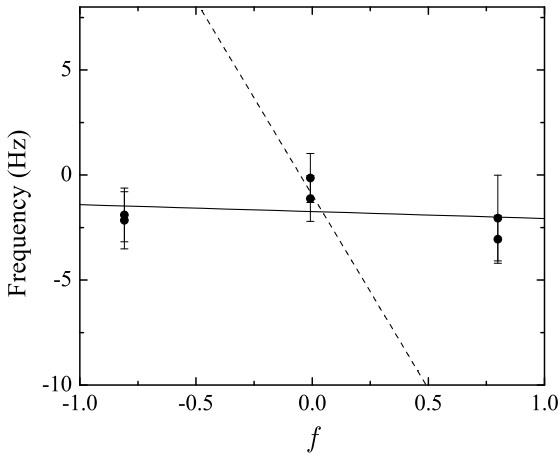}
\caption{\label{fig:fo2}  Testing the exchange symmetry between
the $|1\rangle$ and $|2\rangle$ states.  The transition frequency
is measured as $f$ is varied for a normal cloud at fixed peak
density of $7\times10^{12}$ cm$^{-3}$ and temperature of 510 nK.
The solid line is the fit, which yields
$\alpha^{nc}_{12}/\alpha^{nc}_{11,22}=1.01(2)$, which is to say,
inter- and intra-state density correlations are quite accurately
the same.  The dotted line indicates the expected slope for
$\alpha^{nc}_{12}/\alpha^{nc}_{11,22}=1/2$.}
\end{figure}

\section{\bf Inter-state density correlations}

Exchange symmetry between the $|1\rangle$ and $|2\rangle$ states
can be tested by working at a fixed density and varying the
relative $|1\rangle$ to $|2\rangle$ population by varying the
length of the first Ramsey pulse \cite{ExchangePulse}.  In this
case the first term in Eq.~(4) will be constant and the
measurement will test $\alpha^{nc}_{12}$ and $a_{12}$ as well as
the $\alpha^{nc}_{ii}$ and $a_{ii}$ terms.  To minimize
systematics the interrogation times were kept short, making
precise frequency determination difficult.  Nevertheless, our
measurement (Fig.~\ref{fig:fo2}) indicates
$\alpha^{nc}_{12}/\alpha^{nc}_{11,22}=1.01(2)$, where we have used
the spectroscopically determined scattering lengths.  This clearly
indicates that exchange symmetry is maintained between the
$|1\rangle$ and $|2\rangle$ states.

\begin{figure}
\leavevmode \epsfxsize=3.375in \epsffile{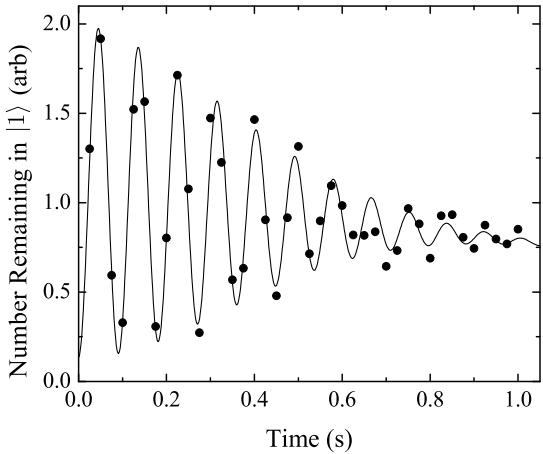}
\caption{\label{fig:chirp}  A data set of Ramsey fringes probing
for frequency shifts as a function of coherence.  For this
measurement normal clouds at a temperature of 480 nK and a peak
density of $3.2\times10^{13}$ cm$^{-3}$ were used.}
\end{figure}

\section{\bf Where's the chirp?}

As a thought experiment, imagine distinct thermal populations of
$|1\rangle$ and $|2\rangle$ atoms, separately prepared, then mixed
together, with the energy of interaction (proportional to
$\alpha^{nc}_{12}$) measured for instance calorimetrically.
Surely in this case the density fluctuations in state $|1\rangle$
and in state $|2\rangle$ would be uncorrelated, and
$\alpha^{nc}_{12}$ would be determined to be 1, not 2.  We lack
the experimental sensitivity to make such a calorimetric
measurement, and our Ramsey-fringe method of measuring energy
differences obviously would not work for incoherent mixtures.  We
speculated, however, that if $\alpha^{nc}_{12}=2$ for coherent
superpositions, and if $\alpha^{nc}_{12}=1$ for incoherent
mixtures, then for partially decohered samples, $\alpha^{nc}_{12}$
would take on some intermediate value.  So by performing a
measurement similar to that in Fig.~\ref{fig:fo2} we might expect
to see a more negative slope for a partially decohered sample;
alternatively a frequency chirp in the Ramsey fringes may be seen
as the sample decoheres.

We probed the time evolution of $\alpha^{nc}_{12}$ in a way
similar to Fig.~\ref{fig:fo2}; however rather than varying $f$ we
set $f\simeq0.8$ then measured $\nu_{12}$ with long interrogation
times, looking for a frequency chirp as the fringe contrast
decreased.  This method has the advantage that there is a
relatively small $|2\rangle$ state population, so effects arising
from $|2\rangle$ loss are minimized.  Seven data sets were taken
for this measurement; an example is shown in Fig.~\ref{fig:chirp}.
By allowing a linear frequency chirp in the fit of the Ramsey
fringes, the frequency shift can be constrained to $-0.2(3)$ Hz by
the time the fringe contrast has reduced to $1/e$
\cite{ChirpCorrect}.  However if we hypothesize that
$\alpha^{nc}_{12}$ goes from $2$ to $1$ linearly as fringe
contrast goes from $100\%$ to $0\%$ we would expect a frequency
shift of $-20(2)$ Hz as the fringe decayed, while the experimental
limit is a factor of 40 smaller.  Clearly this appealing but
unrigorous model is far too naive.

\section{\bf Conclusion}

Where's the chirp?  In truth we don't know.  Our theorist friends
tell us that our confusion arises from our assuming that the
frequency shift $\Delta\nu_{12}$ arise from the difference in
chemical potentials, $\mu_2-\mu_1$.  Instead, they say, we should
directly evaluate the Boltzmann equations for the spin in an
inhomogeneous system \cite{levitov2002,williams2002}. This we have
not as yet done, but even if this solves the mystery in a formal
sense, we are reluctant to give up the cherished notion that the
frequency of the transverse spin precession is a direct
measurement of the energy difference between spin up and spin
down.  For those of us raised in the traditions of atomic physics,
it is a pleasure to note that a two-level system can still yield
surprises, 75 years after the advent of quantum mechanics.

We acknowledge useful conversations with the other members of the
JILA BEC collaboration.  This work was supported by grants from
the NSF and NIST.


\begin{thebibliography}{9}
\bibitem{fspinwave}
At $n=10^{13}$cm $^{-3}$ $\nu_{\mbox{spin wave}}=5$ Hz, so
interrogation times of 20 ms are acceptable.
\bibitem{ExchangePulse}
When we changed the length of the first Ramsey pulse, the second
pulse length was also changed such that the two pulses combined to
form a $\pi$ pulse.
\bibitem{ChirpCorrect}
The quoted chirp limit of -0.2(3) Hz includes a correction to
account for a small frequency change associated with decay of
total atom number during the measurement time.  For each data set
the number decay was found by allowing an exponential decay in the
fit of the Ramsey fringes (not to be confused with the decay in
Ramsey fringe contrast associated with decoherence), resulting in
a correction amounting to -0.74 Hz on average.
\end{thebibliography}
\end{document}